\documentclass{iopart}

\usepackage{natbib}
\usepackage{epsfig}
\usepackage{iopams}

\graphicspath{{/home/cesko/Documents/Lavori/2DLatticeMF/Images}}

\begin{document}

\title[Disordered 2D optical lattices]{Mean-field description of ultracold
  Bosons on disordered two-dimensional optical lattices}
\author{Pierfrancesco Buonsante$^1$, Francesco Massel$^1$, 
        Vittorio Penna$^1$, Alessandro Vezzani$^2$}
\address{$^1$Dipartimento di Fisica, Torino Politecnico,
             Corso Duca degli Abruzzi 24, I-10129 Torino, Italy}
\address{$^2$Dipartimento di Fisica, Universit\`a  degli Studi di Parma and
             C.N.R.-I.N.F.M., 
             Parco Area delle Scienze 7/a, I-43100 Parma, Italy}
\begin{abstract}
  In the present paper we describe the properties induced by disorder on an
  ultracold gas of Bosonic atoms loaded into a two-dimensional optical lattice
  with global confinement ensured by a parabolic potential. Our analysis is
  centered on the spatial distribution of the various phases, focusing
  particularly on the superfluid properties of the system as a function of
  external parameters and disorder amplitude.  In particular, it is shown
  how disorder can suppress superfluidity, while partially preserving the
  system coherence.
\end{abstract}
\pacs{03.75.Lm, 05.30.Jp, 64.60.Cn}
\submitto{\JPB - FTC}
\maketitle           
           
In the last years, the experimental results concerning confined
ultracold atoms in optical lattices have attracted much theoretical
interest from many different fields, ranging from condensed-matter
physics to quantum information theory (see e.g.
\cite{A:Jaksch,A:Calarco}). The ability of tuning the fundamental
physical parameters has provided an unrivaled tool to devise specific
physical situations which, beyond their intrinsic physical interest,
can be employed as important simulation tools \cite{A:Jaksch}. In
particular, the possibility to engineer a defect-free periodic
potential appears as one of the most intriguing characteristics of
these systems. Such ``cleanness'' has allowed the experimental
realization of the Bose-Hubbard (BH) model and the observation of the
superfluid-Mott quantum phase transition discussed by Fisher in his
seminal paper \cite{A:Fisher}, where, in addition, the issue of the
effect of disorder on the physical properties of the BH Hamiltonian
was first addressed. Furthermore many experimental techniques such as
laser speckle field \cite{A:Lye} and the superposition of different
optical lattices with incommensurate lattice constants
\cite{A:Roth03,A:Damski2003,A:Fallani} have recently substantiated the
theoretical investigation of ultracold atoms in disordered optical
lattices \cite{A:Krutitsky,A:Yukalov}, revealing a rich scenario of
new physical situations such as the appearance of new phases (e.g.
Bose-glass phase \cite{A:Damski2003,A:Fallani} ) and superfluid (SF)
percolation in $d$-dimensional ($d>1$) lattices
\cite{A:Sheshadri1995,A:Ospelkaus}.
                        
In this paper we will deal with the properties of a two-dimensional (2D)
optical lattice where bosons are confined by an overall parabolic potential
and subject to a random potential distribution.

The experimental realization of 2D lattices is discussed in \cite{A:Greiner},
\cite{A:Moritz} and \cite{A:Spielman}. In particular the latter represents a
direct experimental realization of the Hamiltonian discussed in the present
paper, when disorder is absent. A valuable feature of the setup here
considered is the possibility to investigate a wide range of physical
situations and geometries according to different external parameter choices.
While the parabolic potential confines bosons in a disk-like domain, the
interplay of the other external parameters (hopping amplitude and chemical
potential), allows one to realize, for example, Mott phases distributed in
concentric shells, each with different filling, intercalated with SF shells.
The 1D counterpart of this scenario has been thoroughly studied in
\cite{A:Batrouni2002}.

Here we will focus particularly on the superfluid-phase spatial distribution
in the presence of a random potential. We analyze first the situation where a
(single) central disk-like Mott region is surrounded by a thin SF shell, and
then the case when the system is fully SF. 
We show that, in the first case, the quasi-1D SF domain is strongly influenced
by the presence of ``impurities'', leading to the drop of the SF fraction for
a small increase of the disorder amplitude. In the second case, the genuine 2D
geometry of the SF domain leads to the formation of circulating streams which
do not disappear abruptly when the disorder amplitude is increased. In this
respect our setup therefore appears to be an effective tool to investigate the
combined effect of disorder and dimensionality on the SF properties of a
bosonic system.
     
The system considered is modeled by a BH Hamiltonian $H = H_I+H_T$, with
           \begin{equation}
              H_I=\frac{U}{2} \sum_{i=1}^M n_i(n_i-1) 
                  -\sum_{i=1}^M \mu_i\, n_i, \, \, 
              H_T=- T \sum_{i,j}^M a_i^\dag A_{ij} a_{j} + h.c.
          \end{equation}
where $n_i=a_i^\dag a_i$, $a_i^\dag$ ($a_i$) represents the creation
(destruction) operator at site $i$, and the total boson number $N=\sum_i n_i$
is a conserved quantity. The random on-site effective chemical potential
$\mu_i$ is given by
           \begin{equation}
             \label{eq:muL}
             \mu_i=-\Omega_0 (x_i^2+y_i^2)+\Delta_i +\mu
           \end{equation}
accounting both for the noise distribution $\Delta$ ($\Delta_i=\Delta \cdot
\xi_i $ with $\xi_i$ uniformly distributed between $0$ and $1$) and for the
parabolic confinement $\Omega_0$.  The Hamiltonian parameters $U$ and $T$
represent the two-body interaction and the hopping amplitude between
neighboring sites respectively, $M$ is the total number of sites (in our
simulations we have chosen $M=31^2=961$) and the matrix $A_{ij}$ is the
so-called adjacency matrix, which is zero if sites $i$ and $j$ are not
connected, and one otherwise.
              
Our analysis is based on the mean-field decoupling scheme already
exploited by many authors for the analysis of the disordered BH model
\cite{A:Damski2003,A:Sheshadri1995,A:Sheshadri1993}. For an
application of this scheme to a disordered system see
\cite{A:Buonsante06CM}. Despite the well known shortcoming of not
taking into account spatial correlations of quantum origin (see e.g.
the discussion about spatial correlations in \cite{A:Wessel}, and in
particular Fig. 4 therein), this mean-field decoupling scheme has
proven the ability to disclose and qualitatively characterize the
prominent physical features of Bosonic atoms loaded in optical
lattices, expecially in dimension greater than one (for a detailed
discussion on the MF scheme and its application see \cite{A:Buonsante}
and references therein). Note however that we take into account the
spatial dependence arising from the disordered+harmonic local
potential, and hence --- at least partially --- the spatial
correlations thereof. The essence of this approximation is the
replacement of two-boson operators $a^\dagger_i a_{i'}$ with
single-site operators weighted by mean-field parameters $\alpha_i =
\langle \psi | a_i | \psi \rangle$ (where $\alpha_i \in \mathbb{C}$
and $| \psi \rangle$ represents the mean-field ground state). Namely
           \begin{equation}
             \label{eq:MFProc}
             a_i^\dag a_{i'} \approx a_i^\dag \alpha_{i'} + \alpha_i^* a_{i'} -
             \alpha_i^* \alpha_{i'} \quad \textrm{with } i\in
             \left[1,\dots,N\right]         
           \end{equation}
           where $\alpha_i$s have to be determined by self-consistency standard
           procedure.
       
Within this scheme, Hamiltonian $H$ can thus be approximated by $H \simeq
\sum_i {\cal H}_i + h $ where $ h = T\sum_{\ell=1}^M \alpha_i^* A_{i\,\ell}
\alpha_{\ell}$ and parameters $\gamma_i= \sum_{\ell=1}^M A_{i\,\ell}
\alpha_{\ell}$ in the single-site Hamiltonian $ {\cal H}_i = U n_i(n_{i}-1) /2
- \mu_i\, n_i - T (\gamma_i a_i^\dag + \gamma_i^* a_i)$ take into account the
original inter-site coupling.
            
Most of the the fundamental physical properties of the system can be obtained
from the values of $\alpha_i$ calculated from Hamiltonian $H$ alone where $T
\in \mathbb{R}$. However, in order to properly describe the SF properties of
the system it is well known that one has to go through same procedure with the
introduction of the \textit{Peierls factors} \cite{A:Shastry}.  The
Hamiltonian with the addition of the Peierls factors reads
       \begin{equation}
         \label{eq:BHP}
           \tilde{H}  = H_I+\tilde{H}_T, \quad 
           \tilde{H}_T=- T \sum_{i,j} \e^{-i \theta B_{ij}} a_i^\dag
           A_{ij} a_{j} + h.c. \,.
       \end{equation}
       
The quantity $\theta B_{ij}$ accounts for an infinitesimal velocity field
$\mathbf{v}$ which can be considered as a ``probe'' to test the superfluidity
properties of the system through the relation $\theta
B_{ij}=m/\hbar\int_{\mathbf{r}_i}^{\mathbf{r}_j} \mathbf{v} \cdot d\mathbf{r}$
\cite{A:Wu} where $\mathbf{r}_i$ is the $i$-th site position. With the Peierls
factors $\tilde{H}$ becomes the Hamiltonian in the moving reference frame
which is analogous to the Hamiltonian of a charged particle in presence of a
magnetic field (see e.g.  \cite{A:Wu}).
For the determination of the response of the system, it is necessary to take
the limit $\theta \to 0$ and hence it is appropriate to consider the presence
of the Peierls factors in perturbative terms with respect to the Hamiltonian
$H$. Up to first order, it is possible to show that the mean-field parameters
for the Hamiltonian $\tilde{\mathcal{H}}$, corresponding through the
mean-field site-decoupling approach to the Hamiltonian $\tilde{H}$, assume the
following form $\alpha_i=\alpha^0_i \exp\left[-i \theta \phi_i \right]$, where
$\alpha_i^0$ represent the (real) mean-field parameter related to the
Hamiltonian $\mathcal{H}$, and the form of the phase term depends on
the first order approximation in $\theta$.
                 
In this paper we define the SF fraction (SFF) as the ratio between the current
operator, as evaluated in Eq.  (\ref{eq:ExpJ}), and the maximum possible SF
current between two sites, i.e.
       \begin{equation}
         \label{eq:sf_frac}
         f_{\rm s}=\frac{\sum_{ij} \langle \psi | J_{ij} | \psi
           \rangle}
         {\sum_{ij}\langle \psi | J^{max}_{ij} | \psi \rangle}
         \quad \textrm{with }\quad 
         \langle \psi | J^{max}_{ij} | \psi \rangle
         =-2T\sqrt{n_i}\sqrt{n_j}B_{ij}
       \end{equation}
          where we have employed the current operator defined as
       \begin{equation}
         \label{eq:CurrOp}
         J_{ij}=i T\left(\exp[-i\theta B_{ij}]a_j^\dag a_i-\exp[i\theta B_{ij}]a_i^\dag a_j \right)
       \end{equation}
which describes the supercurrent in the rotating (lattice) frame.  Recalling
Eq. (\ref{eq:MFProc}), the first-order approximation for the expectation value
of the current operator can be written in the mean-field scheme as
       \begin{equation}
         \label{eq:ExpJ}
       \mathcal{J}_{ij}=\langle \psi | J_{ij} | \psi \rangle
       \simeq -2T\alpha^0_i\alpha^0_j \theta \left[\phi_j-\phi_i+B_{ij}\right].
       \end{equation}
       
\begin{figure}[!h]
         \centering \epsfig{figure=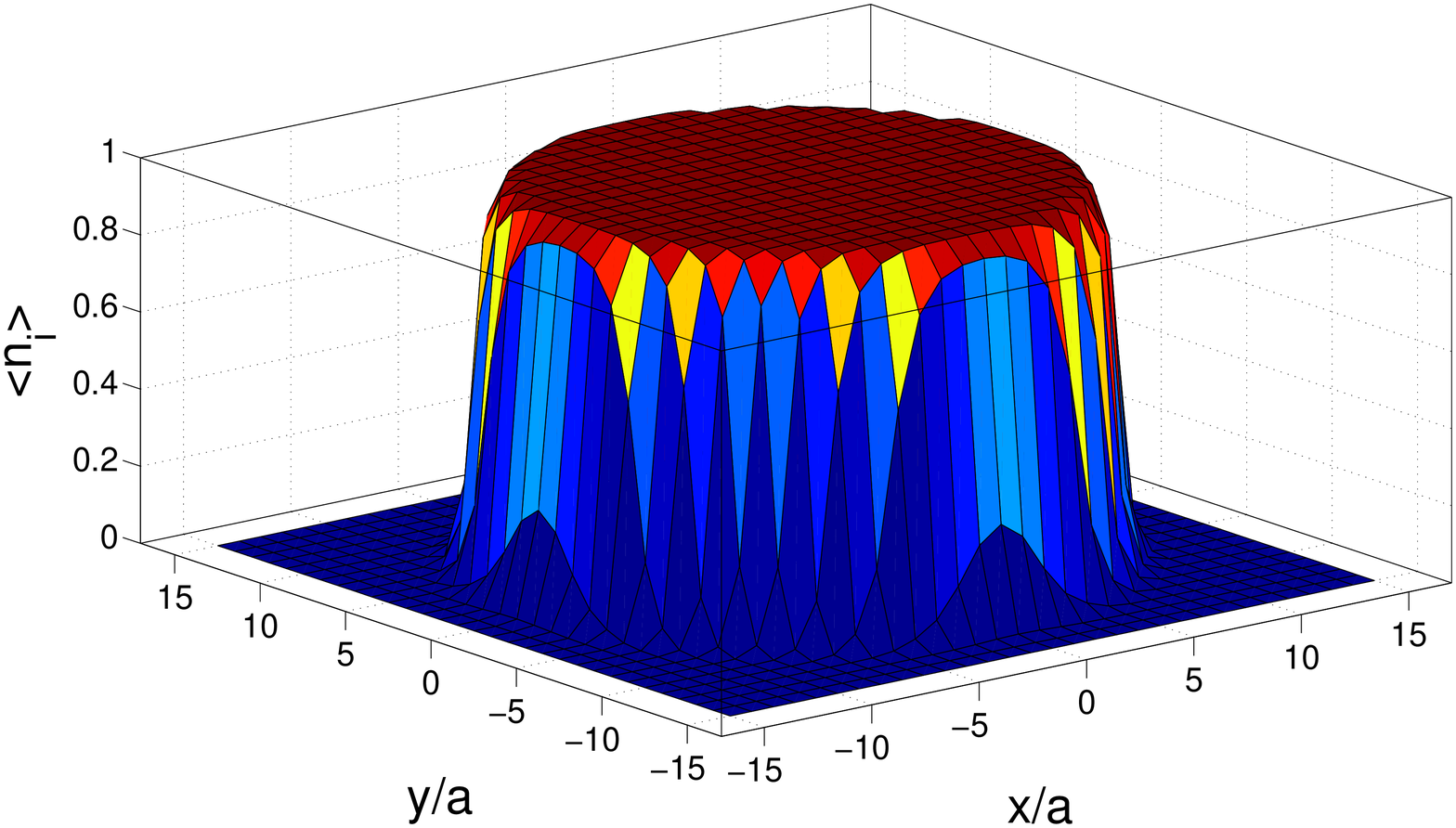,width=0.45\textwidth}
                    \epsfig{figure=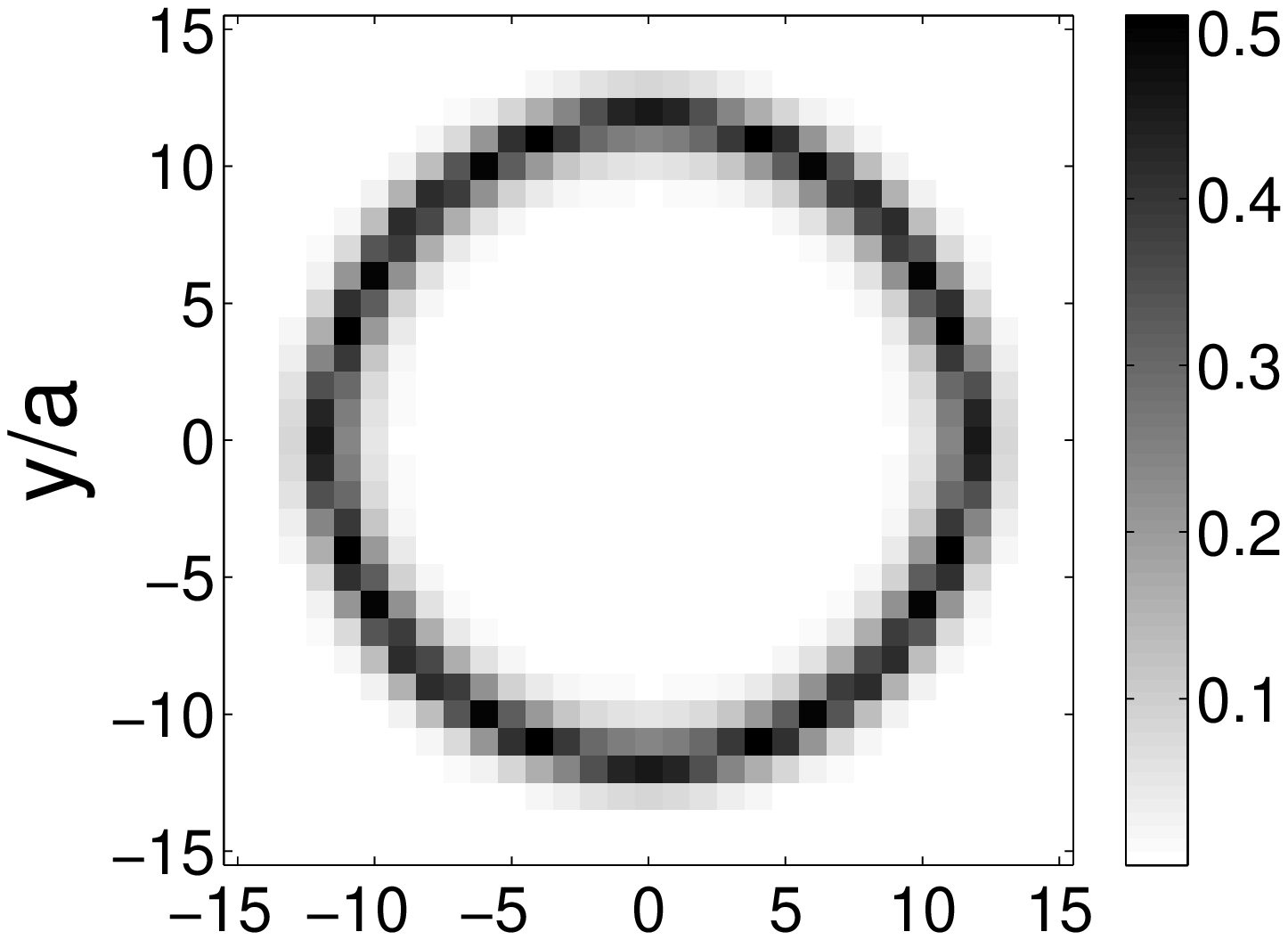,width=0.45\textwidth}\\
         \centering \epsfig{figure=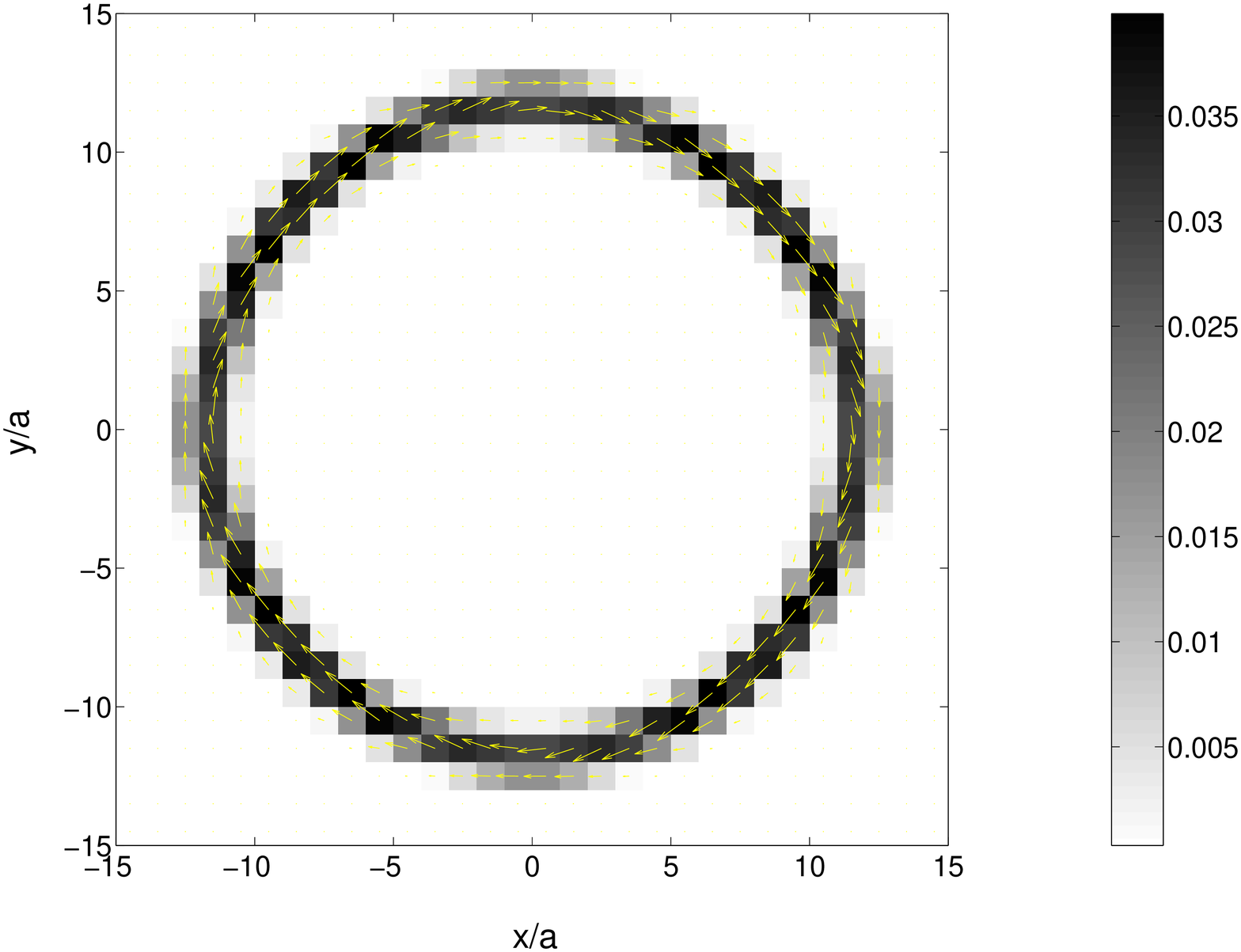,width=0.7\textwidth}
          \caption{\textbf{Upper Panels.} Expectation value for the local number
                    operator (left) and value of $\alpha_i$ (right).
                   \textbf{Lower Panel.} Expectation value for the current
                   operator $J_{ij}$. Both configurations are considered in
                   absence of disorder}
         \label{fig:NoNoise}
       \end{figure}

\begin{figure}[!h]
         \centering 
         \centering \epsfig{figure=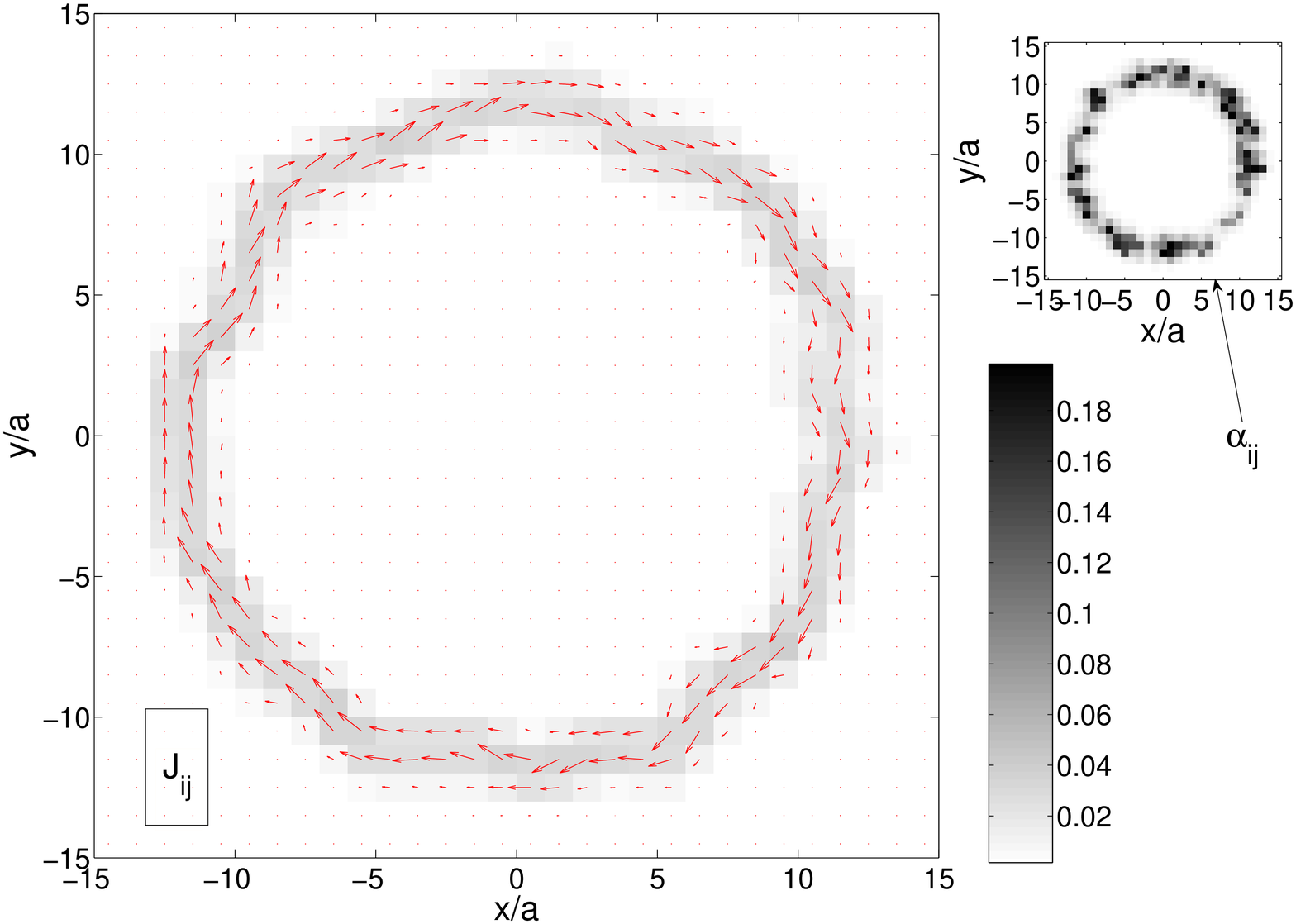,width=0.7\textwidth}
                    \epsfig{figure=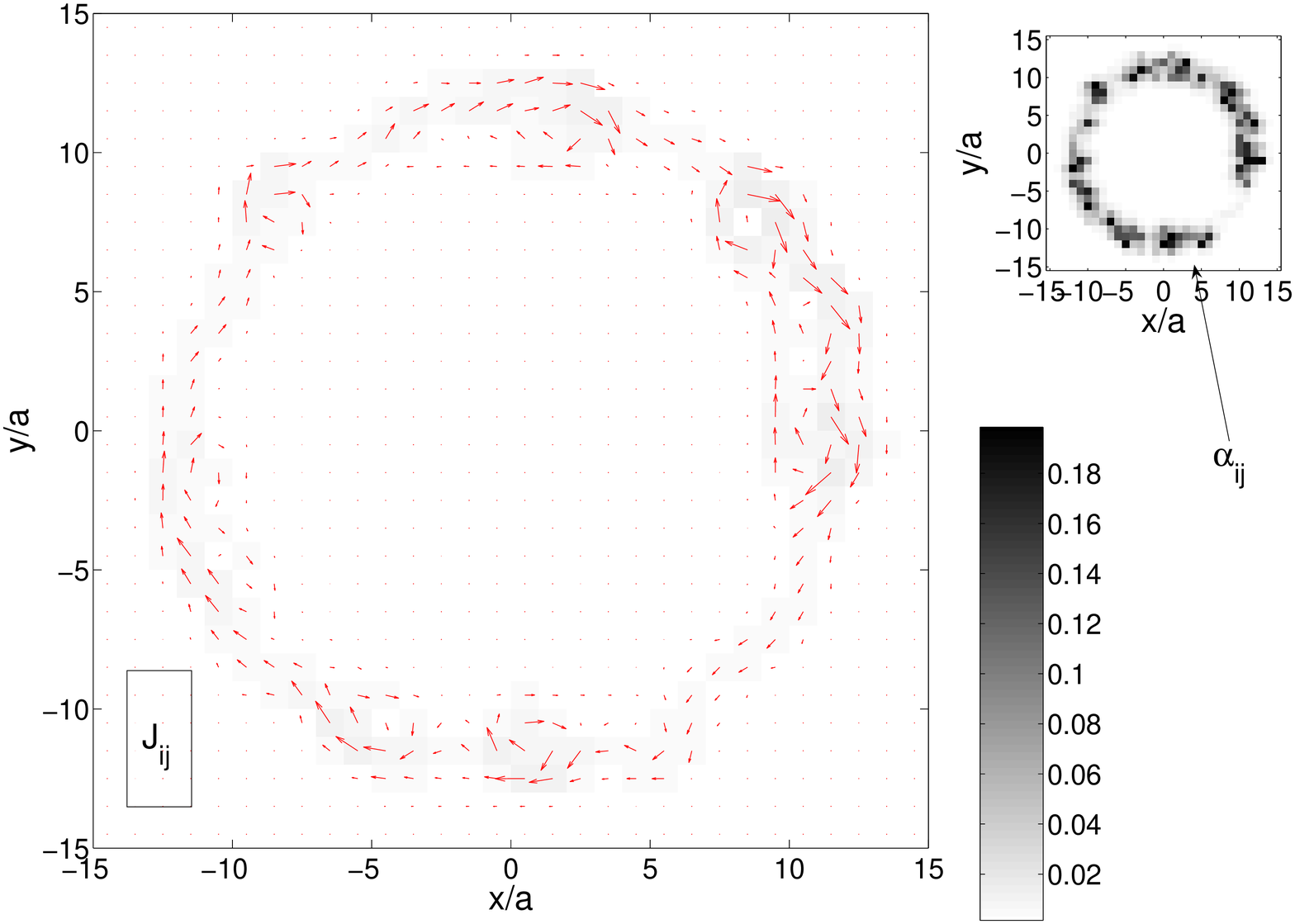,width=0.7\textwidth}
          \caption{Effect of increasing noise amplitude on
            the flux $J_{ij}$ (arb. units) distribution (left) and on the
            $\alpha_i$ distribution (small panel, right). The figures
            considered represent noise amplitudes $\Delta/\Omega_0=$ 60 (top),
            67 (bottom).}
         \label{fig:Noise}
       \end{figure}

So far, we have not explictly defined the structure of the site-dependent
(through matrix $B_{ij}$) Peierls factors. As suggested by the circular
symmetry of the parabolic potential, we have considered by definition $B_{ij}$
an infinitesimal rotation around the paraboloid axis.
The generated flow --observed in the rotating frame-- must be construed as a
response to this infinitesimal rotation. To interpret correctly Figs.
\ref{fig:NoNoise}, \ref{fig:Noise}, and \ref{fig:NoiseHi}, where the spatial
distribution of $\mathcal{J}_{ij}$ is depicted, it necessary to recall that
supercurrents are depicted in the (rotating) lattice frame and hence, in the
laboratory frame, the overall rigid-body rotation effect must be subtracted.
The SFF at rest appears thus to be surrounded by a rotating rigid-body -- in
this case the non-SF fraction.
      
In the 1D homogeneous case $f_{s}$ vanishes iff $\alpha_i=0$ (the standard
condition for Mott regime). In this case, the translational invariance of the
system holds, entailing $n_i=n_j=n$, $\alpha^0_i=\alpha^0_j=\alpha_0$ and
$\phi_i=\phi_j$ (overall phase coherence), thus $\mathcal{J}_{ij}=0$ iff
$\alpha_0=0$. In 2D disordered case, conversely, Eq.  \ref{eq:ExpJ} shows that
the condition $f_{s}=0$ is ensured either by $\alpha_i=0$, like in the 1D
homogeneous case, or, more interestingly, by means of a phase rearrangement,
which is determined by the constraint $\phi_j-\phi_i+B_{ij}=0$, entailing the
emergence of local clusters with $\alpha_i \neq 0$.
       
It is worth noticing that, for a uniform ($\mu_i=\mu$) 1D system, due to translational
invariance, Eq.  (\ref{eq:sf_frac}) reproduces the mean-field behavior of SFF
as defined  in the literature \cite{A:Roth03b}
       \begin{equation}
         \label{eq:sf_frac1D}
         f_{\rm s,1D}=
          \frac{\sum_{ij} 2 T \alpha_0^2 B_{ij}}{\sum_{ij} 2 T n B_{ij}}=
          \frac{\alpha_0^2}{n}.
       \end{equation}
The identification of $f_{\rm s,1D}$ with $\alpha_0^2/ n$ deserves some
comments. In \cite{A:Penrose} it was shown that the largest eigenvalue of the
one-body density matrix $\rho_{ij}=\langle \psi | a^\dagger_i a_j | \psi
\rangle /N$ is a measure of the condensate fraction $f_c$.  Within the
mean-field scheme, it is possible to prove that the ratio ${\langle \alpha_i^2
\rangle} /{ \langle n_i \rangle }$-- the space average ${\langle x_i \rangle}$
($x=\alpha,\, n$) takes into account the lattice inhomogeneity due to the
disorder--, can be related to the largest eigenvalue of the mean-field
approximation of matrix $\rho_{ij}=\left[\left(\langle \psi | n_i | \psi
\rangle - |\alpha_i|^2 \right) \delta_{ij} + \alpha^*_i \alpha_j\right]|/N$
and thus provides an estimate of the condensate fraction $f_c$. In this
spirit, for a uniform system, Eq.  (\ref{eq:sf_frac1D}) satisfies the
condition $f_c \sim f_s$. From a more general point of view, since the
condition $f_c \neq 0$ implies long range correlations, ${\langle \alpha_i^2
\rangle}/{\langle n_i \rangle}$ can be viewed as a good coherence measure even
for a disordered system.
                              
In the numerical simulations we have analyzed two different situations.  In
the first case, we have evaluated the effect of noise on a configuration where
a single central Mott domain, with $n_i=1$, is surrounded by a SF shell (Fig.
\ref{fig:NoNoise}), exhibiting a ring supercurrent. The parameters choice in
this case has been performed in order to obtain a quasi-1D ring-shaped domain.
In the laboratory frame the latter corresponds to inert matter decoupled from
the lattice rotation owing to its SF character. This situation was explored in
\cite{A:Wessel}, in absence of disorder\cite{Nota}. In comparison, as a second
case, we consider a situation where a central disk-like SF region is present
and, consequently, exhibits a 2D behavior.
             
      \paragraph{Case I.}
      
In the first case ($T/U=1.6\, 10^{-2}$, $\Omega_0/U=6\,10^{-3}$ and $\mu/U=0.5$
corresponding to $N_{tot}=199$ total particles) the increase of the noise
amplitude leads to the ``fracture'' of the ring with $\alpha_i \neq 0$ and to
the drop of the SFF, leading the rearrangement of the SF flow within the new
potential pattern (Fig.  \ref{fig:Noise}).
       
In the idealized situation of a perfect 1D ring domain, the presence of an
impurity cutting the ring is sufficient to determine a sudden drop to zero in
the supercurrent induced by the velocity field. In the situation here
depicted, due to the finite radial extent of the SF region such an impurity
occupies an extended domain. In addition to this effect several vortices
emerge which correspond to feeble supercurrents pinned around the sites with
large (disorder-induced) local chemical potential $\mu_i$ (see
Fig.\ref{fig:Noise}, lower panel).
It is worth noticing that in the laboratory frame, the extended impurity
rotates together with the pinned vortices dragged by the lattice.
       
The flow rearrangement in presence of disorder seems to be driven by the
competition between the two terms in Eq. (\ref{eq:ExpJ}).  In a situation
where disorder is weak, the term $B_{ij}$ prevails, leading to a
quasi-proportionality between the forcing term and the SF flow. On the other
hand, in presence of large disorder, the phase pattern is adjusted so as to
satisfy the condition $\phi_j-\phi_i\simeq -B_{ij}$.  In this case the
parameter $\langle\alpha_i^2\rangle/\langle n_i\rangle$, a good SFF measure
for a homogeneous 1D system, while still being a measure of the local
coherence of the system, can not be interpreted anymore as a measure of the
SFF, since the presence of small clusters with $\alpha_i \neq 0$ will
contribute to the overall $\langle\alpha_i^2\rangle/\langle n_i\rangle$ value
but not to the SFF. Notice that here by ``small'' we refer to a three-site
cluster, in that, in a 2D square lattice 4 sites are needed to close a loop
and thus to support a current.

\begin{figure}[!h]
         \centering \epsfig{figure=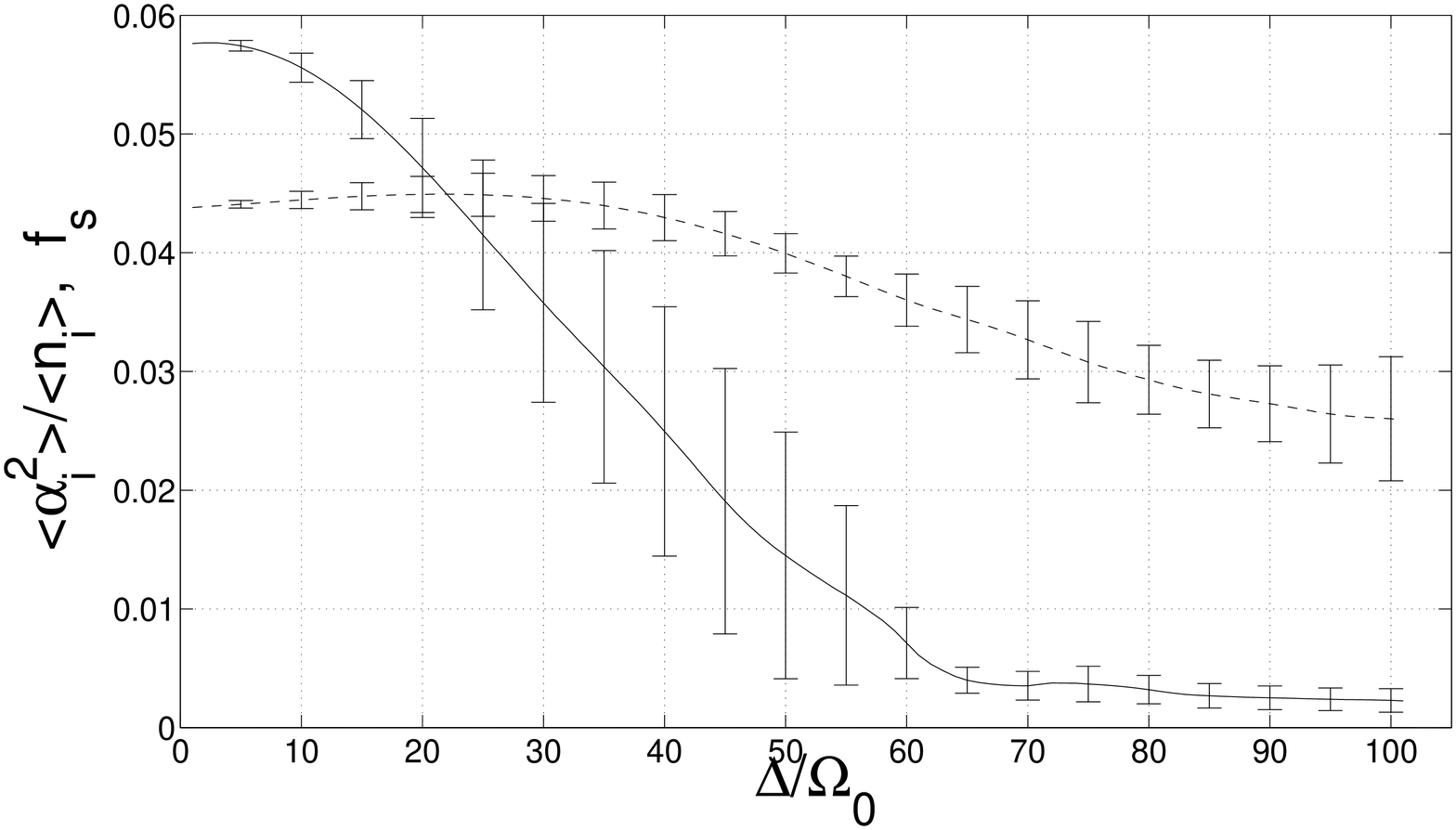,width=0.5\textwidth}
           \caption{Comparison between the SFF
             (solid line) and the value of
             $\langle\alpha_i^2\rangle/\langle n_i\rangle$ (dashed
             line) as a function of disorder amplitude. Error bars account
             for the dispersion
             of the numerical data around their mean value for 10 different
             disorder realizations.}
         \label{fig:sfVsaLo}
       \end{figure}

\begin{figure}[!h]
         \centering \epsfig{figure=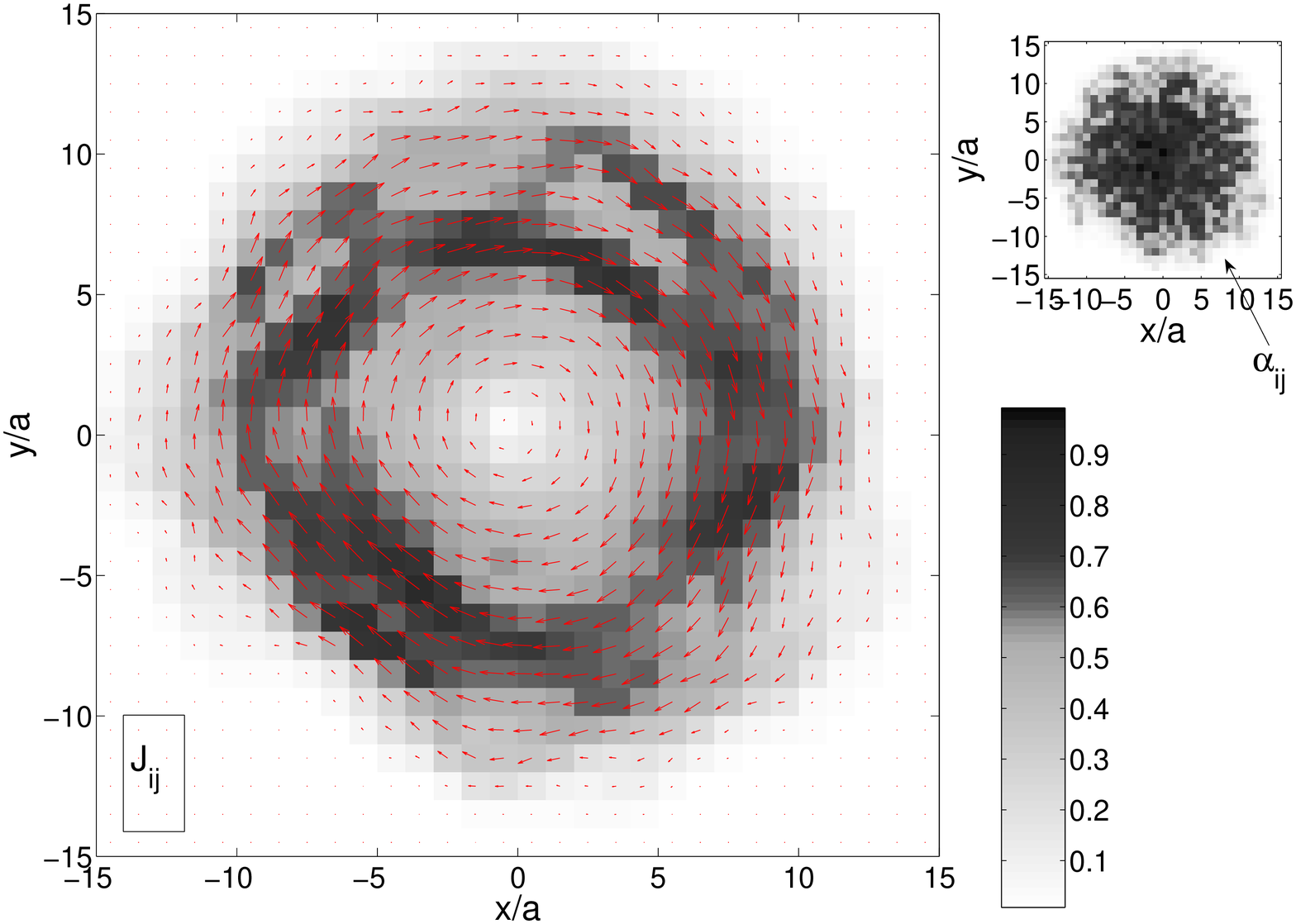,width=0.7\textwidth}
                    \epsfig{figure=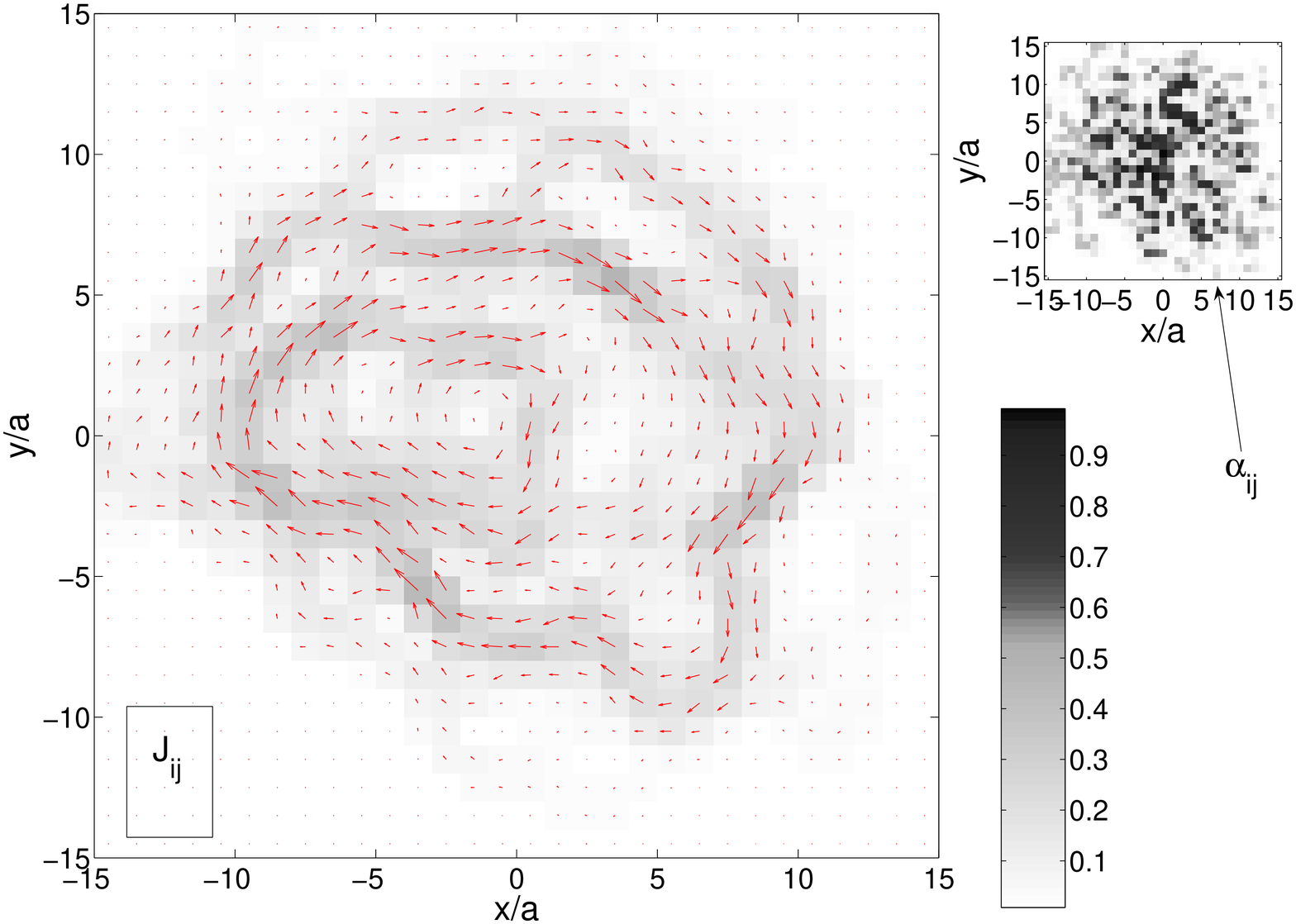,width=0.7\textwidth}
         \caption{Effect of increasing noise amplitude on
            the flux $J_{ij}$ (arb. units) distribution (left) and on the $\alpha_i$
            distribution (small panel, right). The figures considered
            represent disorder amplitudes $\Delta/\Omega_0=$ 200 (top), 600 
            (bottom).}
         \label{fig:NoiseHi}
       \end{figure}

Fig.  \ref{fig:sfVsaLo} shows that an increase of the disorder strength
induces a sharp drop in the SFF while leaving the quantum coherence of the
system almost unchanged.  Each data point is the result of the average over 10
realizations of the random potential. For sake of clarity, the errorbars are
plotted every fifth data point. As we discussed above, the sharp drop in the
SFF corresponds to the interruption of the quasi 1D SF domain characterizing
the system in the absence of disorder
       
A qualitative argument for the determination of the disorder-amplitude range
in which the ring starts to break is to consider that, in the absence of
disorder, the ring is approximately 2-sites thick and with a radius of about
12 sites. Hence the potential energy difference between radially adjacent
sites is about $\Omega_0 \cdot \left(12^2-11^2\right)$. In order to have
superfluidity, the hopping must be large enough to overcome the potential
difference $\Delta_c \simeq \Omega_0 \cdot \left(12^2-11^2\right)$, where
$\Delta_c$ can thus be viewed as the disorder amplitude corresponding to the
drop in the SFF.  As a consequence, if disorder-induced local potentials of up
to the same amplitude are present, the SFF is not affected.
       
       \paragraph{Case II.}
       
       The second setup has been conceived in order to analyze the
       effects of disorder in a fully 2D SF domain. In this case, due
       to the more markedly SF character of the system, it has been
       necessary to consider much larger noise amplitudes to have
       observable effects on the SF distribution. In Fig.
       \ref{fig:NoiseHi} we have sketched the values for
       $\mathcal{J}_{ij}$ and $\alpha_i$ in a situation exhibiting a
       strong SF character ($T/U=0.1$, $\Omega_0/U=3.6\,10^{-2}$,
       $\mu/U=0.6$ with $N_{tot}=313$ ) for increasing values of the
       noise amplitude. Low values for $\mathcal{J}_{ij}$ on the
       lattice center are direct consequence of the velocity pattern
       imposed on the lattice and must not be interpreted as the
       absence of superfluidity at the center of the trap.  As
       previously discussed, in this case we have $\mathcal{J}_{ij}
       \propto -B_{ij}$.
        
\begin{figure}[!h]
         \centering \epsfig{figure=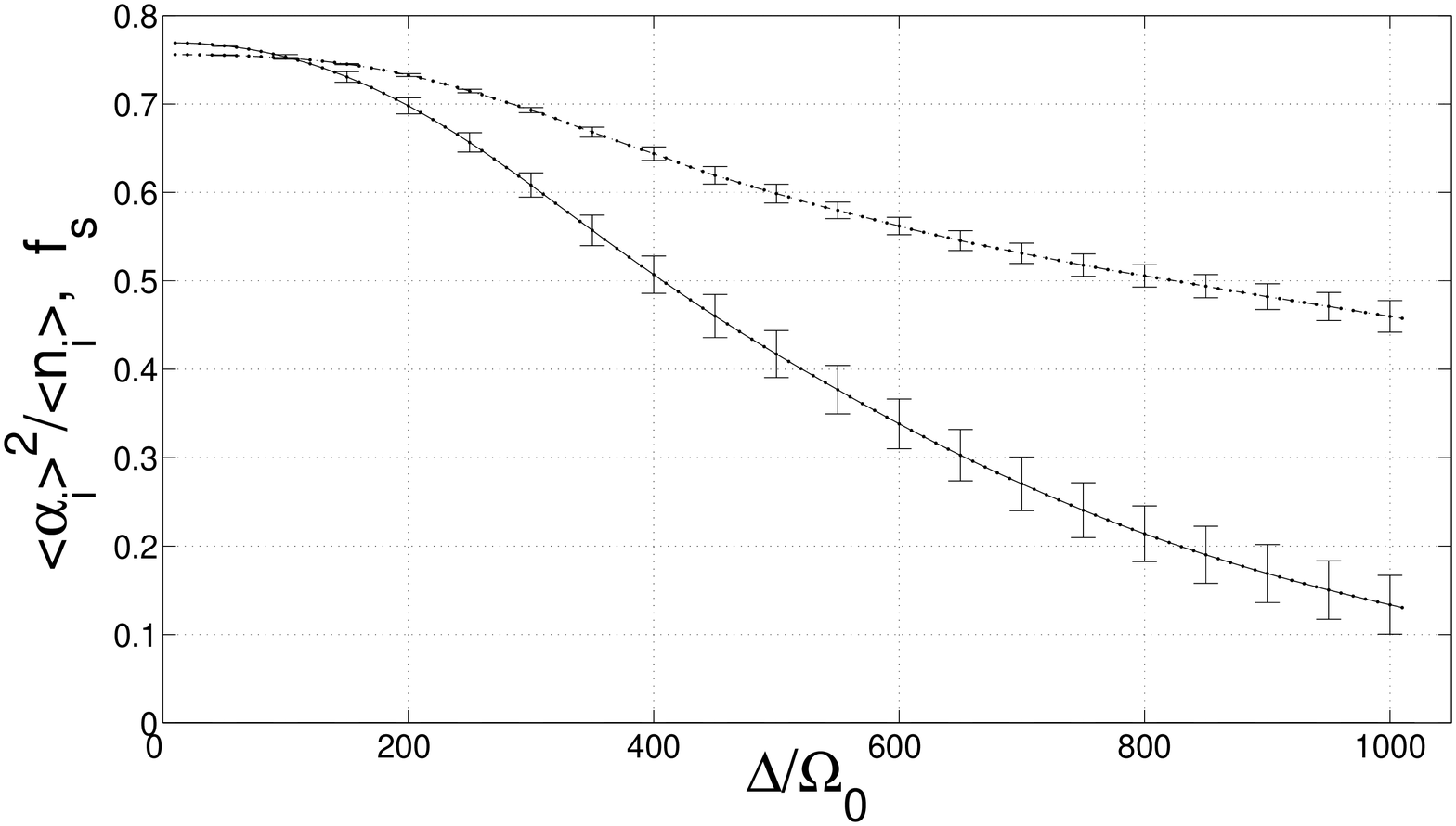,width=0.5\textwidth}
           \caption{Comparison between the SFF (continuous
             line) and the value of $\langle\alpha_i^2\rangle/\langle n_i\rangle$ (dashed line) as a
             function of noise amplitude. Error bars account for the dispersion
             of the numerical data around their mean value for 10 different
             disorder realizations.}
         \label{fig:sfVsaHi}
       \end{figure}

For increasing values of the ratio between the noise and the parabola
amplitude ( $\Delta/\Omega_0$ ), as in \textit{Case I}, the SFF, due to the
rearrangement of the flow, has a slower decrease than in the previous
situation, due to the much larger SFF -- note the scale difference between
Fig.  \ref{fig:sfVsaLo} and Fig. \ref{fig:sfVsaHi}.
      
A prominent difference between \textit{Case I} and \textit{Case II} resides in
the fact that the first one can be considered a quasi-1D case, and hence the
``cutting'' of the SF shell is analogous to the interruption of a 1D chain.
The second case represents, on the other hand, a genuine 2D situation and, as
it is possible to see in Fig.  \ref{fig:NoiseHi}, the presence of noise leads
to the formation of percolation patterns with nonzero circulation
\cite{A:Sheshadri1995}.  The supercurrents in this case will follow closed and
possibly merging paths encircling domains characterized by $f_s=0$ but
$\langle\alpha_i^2\rangle/\langle n_i\rangle \neq 0$. The latter shows then
the survival of a local coherence character, connected to $f_c$, even in the
absence of supercurrents. The condition $f_c \neq 0$ together with $f_s=0$,
according to the scheme of Ref.  \cite{A:Buonsante06CM}, supplies, in the
thermodynamical limit, a possible characterization of a Bose-glass phase.
         
In conclusion, we have presented a theoretical study of the behavior of
Bosonic ultracold atoms in a 2D optical lattice with parabolic radial
confinement with disorder. Our investigation has focused on the SF response
both in the weakly and in the strongly SF regime.  Within our approximation
scheme, we have shown that, in the 2D disordered case, the absence of
supercurrents is determined by two independent effects: the (usual) vanishing
of the parameters $\alpha_i$, and the phase rearrangement implied by
$\phi_j-\phi_i+B_{ij}=0$. While the first effect can be related to the
presence of a Mott phase in the thermodynamical limit, the second one, in the
same limit, suggests the presence of a Bose-glass phase.  In this framework,
we have given a possible interpretation of the mechanism through which
superfluidity is destroyed by disorder. Our numerical analysis evidences how,
as a consequence of the phase rearrangement, the quasi-1D SF domain is
destroyed by the appearance of ``impurities'', with an ensuing drop of the SF
fraction for a small increase of the disorder amplitude.  Conversely, in
\textit{Case II} the genuine 2D nature of the SF domain leads to a
multiple-stream SF flow determined by the increase of the disorder amplitude,
with the consequent appearance of percolation patterns. In the laboratory
frame, the strong-disorder sources in the rotating lattice have an
overall site-dependent dragging effect on lattice bosons indicating the
presence of non-superfluid regions locally surrounded by flows of nonzero
circulation.
       
The rich phenomenology here presented exhibits a complex interplay between
parameters such as the hopping amplitude, the boson-boson interaction, the
disorder amplitude and the parabola coefficient. In view of the intrinsic
interest from the experimental point of view, this deserves a more systematic
analysis we will perform elsewhere.

\def\newblock{\hskip .11em plus .33em minus .07em}
\bibliographystyle{unsrt}

\begin{thebibliography}{10}
\bibitem{Nota}
In order to corroborate the general validity of our method, we notice that the
mean-field results in absence of disorder described in Fig. \ref{fig:NoNoise}
seem to properly reproduce the main findings of the Quantum Mone Carlo
approach of \cite{A:Wessel}.
\end{thebibliography}

\end{document}